\documentclass{article}

\usepackage[nonatbib,final]{neurips_2020}

\usepackage[utf8]{inputenc} 
\usepackage[T1]{fontenc}    
\usepackage{hyperref}       
\usepackage{url}            
\usepackage{booktabs}       
\usepackage{amsfonts}       
\usepackage{nicefrac}       
\usepackage{microtype}      
\usepackage{graphicx}
\usepackage{subfig}
\usepackage{biblatex}
\usepackage{multicol,multirow,booktabs}

\title{%
  Mapping fNIRS to fMRI with Neural Data Augmentation and Machine Learning Models
}

\author{%
  Jihyun Hur \\
  Department of Psychology \\
  Seoul National University, Korea \\
  \texttt{hur\_jihyun@snu.ac.kr} \\
  \And
  Jaeyeong Yang \\
  Department of Psychology \\
  Seoul National University, Korea \\
  \texttt{urisa12@snu.ac.kr} \\
  \AND
  Hoyoung Doh \\
  Department of Psychology \\
  Seoul National University, Korea \\
  \texttt{comicroad11@snu.ac.kr} \\
  \And
  Woo-Young Ahn \\
  Department of Psychology \\
  Seoul National University, Korea \\
  \texttt{wahn55@snu.ac.kr} \\
}

\begin{document}

\maketitle

\begin{abstract} 
Advances in neuroimaging techniques have provided us novel insights into understanding how the human mind works. Functional magnetic resonance imaging (fMRI) is the most popular and widely used neuroimaging technique, and there is growing interest in fMRI-based markers of individual differences. However, its utility is often limited due to its high cost and difficulty acquiring from specific populations, including children and infants. Surrogate markers, or neural correlates of fMRI markers, would have important practical implications, but we have few stand-alone predictors for the fMRI markers. Here, using machine learning (ML) models and data augmentation, we predicted well-validated fMRI markers of human cognition from multivariate patterns of functional near-infrared spectroscopy (fNIRS), a portable and relatively inexpensive optical neuroimaging technique. We recruited 50 human participants who performed two cognitive tasks (stop signal task and probabilistic reversal learning task), while neural activation was measured with either fNIRS or fMRI at each of the total two visits. Using ML models and data augmentation, we could predict the well-established fMRI markers of response inhibition or prediction error signals from 48-channel fNIRS activation in the prefrontal cortex. These results suggest that fNIRS might offer a surrogate marker of fMRI activation, which would broaden our understanding of various populations, including infants.
\end{abstract}

\section{Introduction}

Neuroimaging technique is a crucial analytic tool to probe neural markers of individual differences in decision-making and learning \cite{math_2009, kable_2015, schultz_1997}. Due to its non-invasive nature and high spatial/temporal resolution, functional magnetic resonance imaging (fMRI) has been extensively used to study human populations in cognitive neuroscience and related fields. Numerous studies using advanced analytical methods have revealed how mental processes and states are represented in the brain and they map onto neural activity \cite{yarkoni2011large, norman2006beyond}. Consequently, there is growing interest in fMRI-based (bio)markers in predicting individual differences and decoding mental states \cite[e.g.,][]{woo2017building, kragel2016decoding}. However, fMRI has several practical constraints because of the MRI environment and its high cost \cite{scarapicchia2017functional}. 

Functional near-infrared spectroscopy (fNIRS) has great portability and tolerance for head motion, and has emerged as a promising alternative neuroimaging technique although fNIRS also has its technical constraints such as limited depth of recording (i.e., limited to measuring cortical activity). While numerous studies have simultaneously recorded fMRI and fNIRS \cite[e.g.,][]{schroeter2006investigating, steinbrink2006illuminating} and examined their relationships to show fNIRS as a potential predictor of fMRI markers, few examined whether fNIRS measurement can be directly mapped onto fMRI signals when obtained separately. Hence, it has been difficult to juxtapose the fNIRS and fMRI outputs from different studies and interpret them together. This hinders our understanding of neurological development as infant research have mostly used fNIRS while many adult studies have exploited fMRI. 

To address the gap, we acquired fNIRS and fMRI data independently while human participants were performing the same cognitive tasks. A neural data augmentation technique \cite{naga_2020} and four machine learning (ML) models (see Methods for more details) were then applied to multivariate fNIRS activation patterns to test if they can predict fMRI markers.  

\section{Methods}


\begin{figure}
\centering
\includegraphics[width=\textwidth]{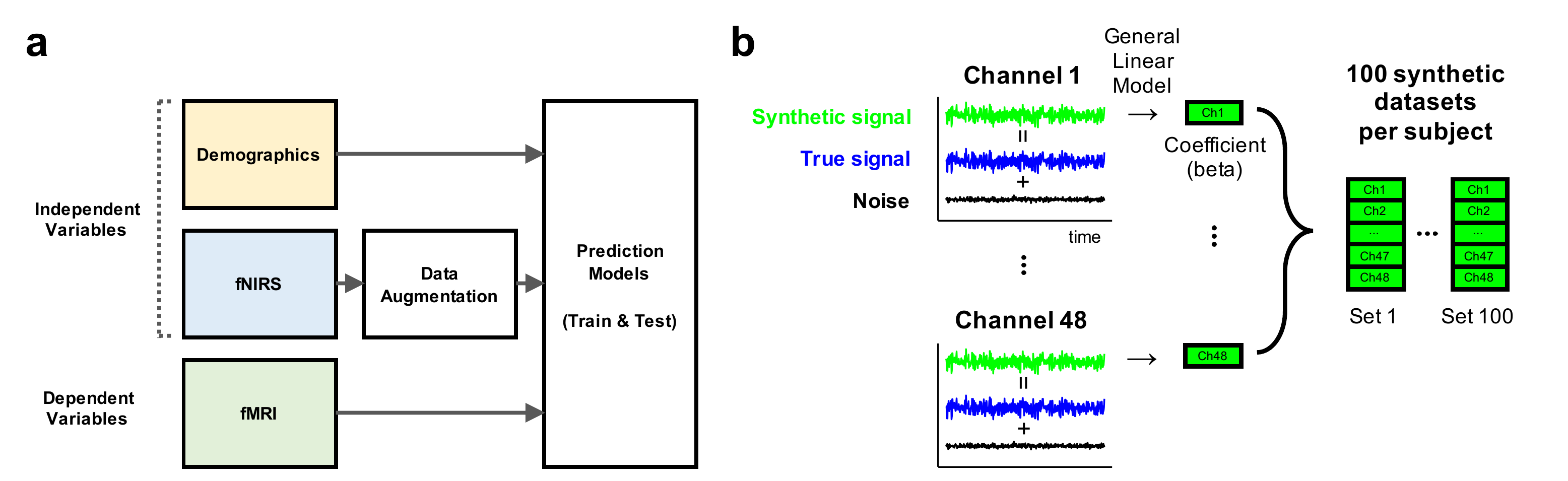}
\caption{(a) Pipeline for fNIRS-fMRI prediction, (b) A graphical illustration of data augmentation.}
\label{fig:methods}
\end{figure}


\subsection{Dataset}

Excluding 2 subjects due to attrition, 48 healthy adults participated in both fNIRS and fMRI sessions that were 2 days apart on average. To examine both low- and high-level cognitive abilities and their neural mechanisms, we asked participants to perform the stop signal task \cite{li_2006} and the probabilistic reversal learning task \cite{ham_2006}. Across two visits, we showed high consistency in task performance by examining high correlation between behavioral measures: stop signal response time for measuring individual difference of response inhibition in the SST, $r = 0.68, p < 0.001$; the number of reversals in the PRL, $r = 0.40, p = 0.01$. 

The Stop Signal Task (SST) is to assess response inhibition, an ability to inhibit actions \cite{li_2006}. The task predominantly requires `Go' actions but occasionally signals to stop the response. To understand the neural correlates of successful response inhibition, we subtracted the `successful go' value from the `successful stop' beta estimates. After data quality control (e.g., head motions, noisy scanner issues), 34 subjects were included in the analysis.

The Probabilistic Reversal Learning Task (PRL) is a reinforcement learning task in which high-level cognitive abilities such as value encoding and prediction error calculation are required \cite{ham_2006}. The task shows two stimuli associated with either probabilistic monetary reward or punishment per trial, and a participant has to make a series of decisions to maximize total reward. We applied hierarchical Bayesian analysis to obtain the trial-by-trial measures of prediction errors for the chosen option using the hBayesDM package in R \cite{ahn_2017}. Then we computed a beta value for the prediction errors computed for all trials per subject. 32 subjects were included in the analysis after quality control.

\subsection{Modalities}

\emph{fNIRS} is a non-invasive optical neuroimaging technique that measures hemodynamic responses in the brain using near-infrared light. The device we used (NIRSIT; OBELAB, Seoul, Korea) is composed of 24 sources and 32 detectors, which configure 48 channels covering the prefrontal cortex. Three measures representing hemodynamic response variation were calculated: oxygenated hemoglobin (HbO), deoxygenated hemoglobin (HbR), and total hemoglobin (HbT). All of them were used for data augmentation. 

\emph{fMRI} has been the most prominent non-invasive neuroimaging technology that records the blood oxygen level dependent (BOLD) signal, intensity of which is determined by the concentration of HbO. We measured the whole-brain activation using a 3T scanner (Magnetom Trio; Siemens, Germany) and included the mean activation of each significantly-activated cluster in the prediction analysis.

For both modalities, we applied the most common neuroimaging analysis method, general linear modeling, to estimate the beta coefficients of event regressors from time-series neural data and convoluted response functions. Then we constructed two main contrasts to analyze. The resultant beta values from fNIRS and fMRI data were used as independent and dependent variables, respectively.

\subsection{Data Augmentation}
Data augmentation is a technique to generate synthetic data by modifying the actual data (e.g., rotation, crop, noise).
In this study, we applied data augmentation to our fNIRS data by adding Gaussian noise.
Each subject's fNIRS data is a $(T_n, 48)$ matrix where $T_n$ indicates the total number of time points of the $n$-th subject and 48 represents the number of channels. We normalized the data by channel and created the 100 same-sized matrices containing Gaussian noise with the mean of 0 and standard deviation of 0.01. By adding each of the noise matrices to the original matrix, we generated 100 artificial dataset per subject (Figure~\ref{fig:methods}b).

\subsection{Prediction}
In the prediction, we used 4 traditional ML models: linear regression, Lasso regression~\cite{lasso}, ridge regression~\cite{ridge}, and support vector regression (SVR)~\cite{svm} with radial basis function kernel. Note that traditional ML and deep learning models often show similar performance when applied to neuroimaging data \cite{schulz2020different}. We trained our models on the synthetic fNIRS beta dataset and fit each estimated model to true fNIRS beta values. To predict the true fMRI beta values, we used the leave-one-out cross-validation procedure and evaluated model performance by measuring the r-squared ($R^2$) values, correlation coefficients, their corresponding p-values, and the mean squared error (MSE).

\section{Results}

\begin{table}[t!]
\small
\centering
\begin{tabular}{llrrrr}
\toprule
& & \multicolumn{3}{c}{SST} & \multicolumn{1}{c}{PRL} \\
\cmidrule(l){3-5}\cmidrule(l){6-6}
& Model & Right IFG & SMA & Left IFG & IPL \\
\midrule
\multirow{4}{*}{HbO}
& Linear reg. & 5.125 & 31.391 & 98.099 & 1.622\\
& Lasso reg. & 7.870 & 16.782 & 13.839 & 0.272\\
& Ridge reg. & 4.747 & 21.245 & 71.106 & 0.530\\
& SVR (RBF) & 7.108 & 19.035 & 16.723 & 0.202\\
\midrule
\multirow{4}{*}{HbR}
& Linear reg. & 20.812 & 31.887 & 44.019 & 0.850\\
& Lasso reg. & \textbf{4.787} & \textbf{7.194} & \textbf{8.158} & 0.265\\
& Ridge reg. & 18.626 & 28.676 & 38.007 & 0.490\\
& SVR (RBF) & 7.515 & 10.710 & 12.522 & 0.277\\
\midrule
\multirow{4}{*}{HbT}
& Linear reg. & 26.669 & 49.740 & 31.752 & 0.379\\
& Lasso reg. & 7.295 & 15.826 & 16.270 & 0.220\\
& Ridge reg. & 20.488 & 41.420 & 27.691 & 0.186 \\
& SVR (RBF) & 8.129 & 15.718 & 14.708 & \textbf{0.115} \\
\bottomrule
\end{tabular}
\caption{%
Comparison of the mean squared error (MSE) across 4 models with different fNIRS signals.
IFG = inferior frontal gyrus.
SMA = supplementary motor area.
IPL = inferior parietal lobule.
}
\label{tbl:pred-result}
\end{table}

\begin{figure}
\centering
\includegraphics[width=\textwidth]{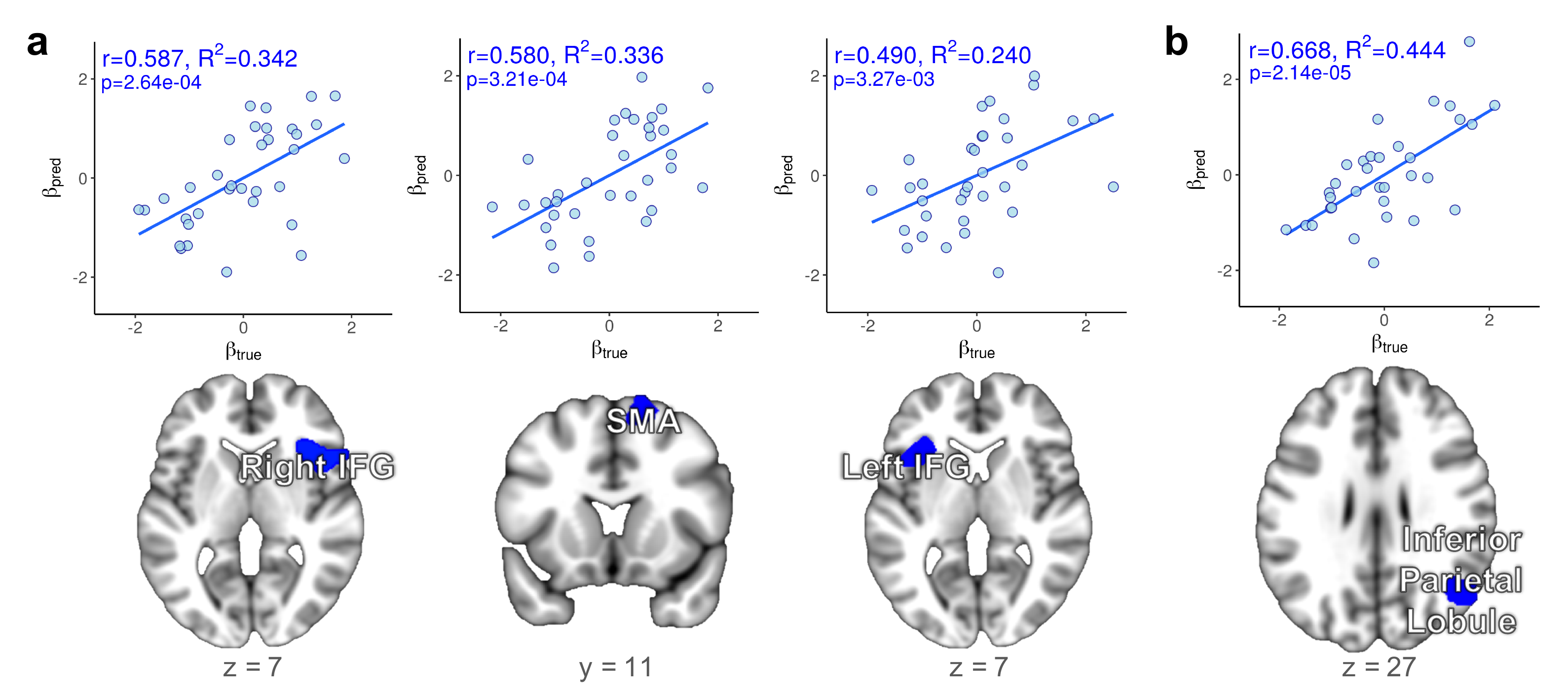}
\caption{Prediction results on the activated brain clusters found in the (a) SST and (b) PRL. The x and y axes of the upper panes show the z-scores of the predicted fMRI mean beta and the actual beta values.}
\label{fig:pred-results}
\end{figure}

\paragraph{SST}
In the fMRI data analysis, 8 clustered regions showed significant activation during successful response inhibition. The Lasso regression model with the HbR signals resulted in the best prediction of 3 out of the 8 clusters (Table~\ref{tbl:pred-result}). Each of the significantly predicted clusters is located in a distinct region in the brain as shown in Figure~\ref{fig:pred-results}a. The three clusters include the right and left inferior frontal gyrus (IFG) in the prefrontal cortex and the supplementary motor area (SMA) in the superior frontal cortex. 

\paragraph{PRL}
In prediction with the reversal learning task data, we identified the brain regions significantly associated with the trial-by-trial prediction error values. We replicated the previous neural findings that the subcortical brain areas including the striatum are engaged in encoding the prediction errors. However, we were not able to predict the activation in the striatum cluster. Instead, the SVR-RBF model with the HbT signals well predicted the mean fMRI activation in the inferior parietal lobule (IPL; Figure~\ref{fig:pred-results}b) located in the cerebral cortex.

\section{Discussion} 

With data augmentation and our prediction pipeline (Figure~\ref{fig:methods}a), we predicted brain areas that engage in response inhibition and encoding of prediction errors. fNIRS beta values from the HbR signals with the Lasso regression model predicted fMRI clusters that are located in the right IFG, SMA, and left IFG. This results are consistent with the previous findings that the fMRI BOLD signals are dependent on and highly associated with fNIRS HbR signals~\cite{strang_2002}. In addition, bilateral IFG have been known to be closely associated with individual differences in inhibitory control~\cite{li_2006}. By predicting the fMRI activation level with fNIRS data, we found that fNIRS can be used to predict fMRI correlates of response inhibition.

We could also predict individual differences in the neural encoding of prediction errors in the PRL. We found that the beta maps extracted from the fNIRS analysis can reliably predict the fMRI activity in the IPL. In previous studies, the IPL represented prediction errors and showed strong functional connectivity with the striatum, the region well-known to encode prediction errors ~\cite{garr_2013, heil_2019, mar_2008}. One limitation, however, is that we could not find a good model to predict the activation in the striatum. This finding is still consistent with the previous findings that fNIRS is better suited to measure and predict neural activation within the cortices, instead of the subcortical brain regions \cite{ferr_2012, niu_2015, plichta_2005}. 

We think one of the main contributions of this study is our prediction pipeline (Figure \ref{fig:methods}). Applying data augmentation directly to neural data is a recent trend \cite{naga_2020, safdar_2020}, and this study further supports that data augmentation is a useful and promising tool in predictive analysis with fNIRS. Despite these promising results, there are some limitations of this study. First, we did not investigate whether we could predict functional connectivity examined through fMRI measurement with fNIRS. Second, confounding variables such as environmental difference or emotional states might have affected the neural activation, which requires further investigation. In summary, this study demonstrated the potential utility of fNIRS as a surrogate measure of fMRI-based markers, which may have great utility in infant brain research.

\begin{ack}
The research was supported by the Institute for Information and Communications Technology Planning and Evaluation (IITP) grant funded by the Korea government (MSIT) (No. 2019-0-01367, BabyMind), the Basic Science Research Program through the National Research Foundation (NRF) of Korea funded by the Ministry of Science, ICT, and Future Planning (NRF-2018R1C1B3007313 and NRF-2018R1A4A1025891), and the Creative-Pioneering Researchers Program through Seoul National University to W.-Y.A.

\end{ack}

\printbibliography

\end{document}